\pdfoutput=1

\documentclass[twocolumn]{aastex62}

\usepackage{epstopdf}
\usepackage{ulem}

\shorttitle{Continuous null-point reconnection}
\shortauthors{Zou et al.}

\begin{document}

\title{Continuous Null-Point Magnetic Reconnection Builds Up a Torus Unstable Magnetic Flux Rope Triggering the X9.3 Flare in Solar Active Region~12673}

\correspondingauthor{*Chaowei Jiang}
\email{chaowei@hit.edu.cn}

\author{Peng Zou}
\affil{Institute of Space Science and Applied Technology, Harbin Institute of Technology, Shenzhen 518055, China}

\author{Chaowei Jiang*}
\affil{Institute of Space Science and Applied Technology, Harbin Institute of Technology, Shenzhen 518055, China}

\author{Fengsi Wei}
\affil{Institute of Space Science and Applied Technology, Harbin Institute of Technology, Shenzhen 518055, China}

\author{Xueshang Feng}
\affil{Institute of Space Science and Applied Technology, Harbin Institute of Technology, Shenzhen 518055, China}

\author{Pingbing Zuo}
\affil{Institute of Space Science and Applied Technology, Harbin Institute of Technology, Shenzhen 518055, China}

\author{Yi Wang}
\affil{Institute of Space Science and Applied Technology, Harbin Institute of Technology, Shenzhen 518055, China}

\begin{abstract}
Two X-class solar flares occurred on 2017 September 6 from active region NOAA 12673: the first one is a confined X2.2 flare, and it is followed only $\sim 3$ hours later by the second one, which is the strongest flare in solar cycle 24, reaching X9.3 class and accompanied with a coronal mass ejection. Why these two X-class flares occurred in the same position with similar magnetic configurations, but one is eruptive while the other is not? Here we track the coronal magnetic field evolution via nonlinear force-free field extrapolations from a time sequence of vector magnetograms with high cadence. A detailed analysis of the magnetic field shows that a magnetic flux rope (MFR) forms and grows gradually before the first flare, and shortly afterwards, the MFR's growth is significantly enhanced with a much faster rise in height, from far below the threshold of torus instability to above it, while the magnetic twist only increases mildly. Combining EUV observations and the magnetic field extrapolation, we found that overlying the MFR is a null-point magnetic topology, where recurrent brightening is seen after the first flare. We thus suggest a scenario to interpret the occurrence of the two flares. The first flare occurred since the MFR reached a high enough height to activate the null point, and its continuous expansion forces the null-point reconnection recurrently. Such reconnection weakens the overlying field, allowing the MFR to rise faster, which eventually crosses the threshold of torus instability and triggers the second, eruptive flare.
\end{abstract}


\section{Introduction}
\label{sec:intro}

Solar flares are the most attractive events on the Sun. They release large amount of magnetic energy and transform them into radiative, thermal and kinetic energy. They can generate solar energetic particles and coronal mass ejections (CMEs) which can threaten the space environment of our Earth directly. Among them, the X-class ones, which mean the extreme ones, are most powerful and most threatening. Therefore, study of their trigger mechanism and eruptiveness is an important topic not only in understanding the underlying physics but also in forecasting the space weather.

It is commonly believed that the energy released in solar flares originates from magnetic energy stored in the solar corona, in particular, in sheared or twisted magnetic flux tubes \citep{forbes2000}. The standard flare model is well known as the so-called CSHKP model \citep{carmichael1964,sturrock1966,hirayama1974,kopp1976}. It proposed that some kinds of instability initially caused the upward motion of a magnetic flux rope (MFR) and eventually lead to the reconnection between the underneath field lines. According to the standard model, flares is often accompanied by an eruption or CME, which was confirmed by lots of observations \citep[e.g., see][]{chen2011}. Nevertheless, the famous active region NOAA AR 12192 shows an unexpected behavior, since it generates 6 flares of X-class and tens of smaller ones but almost none of them is accompanied with CME \citep{sun2015,thalmann2015,liu2016b}. It is unusual to see a confined X-class flare, since previous statistic show that ~90\% X-class flares associated with a CME \citep{yashiro2005,wang2007}. Thus it is interesting to study why an X-class flare is confined. And some analysis was done in recent decades \citep{schmahl1990,feynman1994,gaizauskas1998,green2002,wang2007,liu2008,cheng2011,chen2013,liu2014,zou2019}. It seems that the overlying magnetic field contributes most in constraining an eruption \citep{wang2007,liu2008,cheng2011,chen2013,sun2015,liu2016b,zou2019}. But for an accurate comparison, two X-class flares that occurred with the similar magnetic environment but display different behavior will provide a stronger conclusion in analysing their eruptiveness. Fortunately, some authors reported such events and these two X-class flares occurred not only in the same position but also, interestingly, the same day \citep{liu2019,zou2019}. The active region where they took place is NOAA AR 12673.

The AR 12673 is a very active AR in solar cycle 24. During the time period its passing the solar disc, four X-class flares, 27 M-class flares and numerous smaller flares took place there \citep{moraitis2019}. Especially the X9.3 flare occurred at 11:53 UT in September 6 is the strongest flare of the solar cycle. Many papers have been devoted to study this flare. Observations suggest this flare may be caused by the kink instability of a filament \citep{yang2017}. Investigation of the pre-flare photospheric motions and magnetic properties are also investigated by several authors \citep{yan2018,romano2018,verma2018}. They reported that the high magnetic gradient cross the polarity inversion line (PIL), the fast helicity injection and high ratio of non-potential helicity to total helicity all contributed to the huge eruption of the flare. Basing on coronal magnetic field extrapolations, \citet{hou2018} considered its eruption should be due to the interaction of a multi-MFR system, which is similar to the investigation of \citet{liu2019}. Data-constrained MHD simulation of \citet{jiang2018, inoue2018} reproduced the reconnection process between different twisted magnetic-flux systems, which eventually form a larger and coherent MFR during the eruption. They suggested that the torus instability play an leading role in triggering and driving the eruption.

Only 3 hours before the X9.3 one, namely at 8:57~UT in the same day, another confined X2.2 flare took place in the same location, and moreover these two flares have flare ribbons along the same PIL \citep{jiang2018,zou2019,liu2019}. In our prior study \citep{zou2019}, we selected two vector magnetograms for each flare, one observed immediately before the flares and the other one after the flares, and extrapolated the coronal magnetic field using them for comparing the differences between the two flares. It is found that there is MFR structure before both the two flares, but the key difference between the confined and the eruptive ones is that the MFR in the latter reaching the threshold of torus instability while the former is far below the torus instability. Observation, theoretical and numerical studies have suggested that in the presence of a well-defined MFR, the torus instability is controlled mainly the decay index of the background field overlying the MFR~\citep{aulanier2010,jiang2018,zou2019}. Since the two flares occurred so closely in time, the background overlying magnetic field should not behaves such a huge difference. Thus we suspect that in this short time period the MFR survived in the confined flare rises significantly in height and trigger the torus instability in the eruptive flare. Then, how did the MFR rise such height after the previous flare?

In this paper, we attempt to answer this question. We carried out a time sequence of magnetic field extrapolations to follow evolution of the MFR, which provides new insight in understanding the pre-eruption process and the trigger of the eruption. In the following, we first describe the data and method in Section~\ref{obs}, then present the results in Section~\ref{res}, and finally conclude in Section~\ref{dac}.

\section{Data and Method} \label{obs}

We utilized the time sequence of vector magnetograms and Extreme Ultraviolet (EUV) images of NOAA AR12673 from 00:00~UT to 12:00~UT on 2017 September 6. The EUV images were taken by the Atmospheric Imaging Assembly \citep[AIA,][]{lemen2012} on board the \textit{Solar Dynamics Observatory} \citep[SDO,][]{pesnell2012}. It can gather seven EUV filtergrams which covers the temperature from 10$^5$K to 10$^{7}$K with a spatial resolution of 0.6\arcsec pixel$^{-1}$ and a cadence of 12s. It can observe both the flare ribbons embedded in chromosphere and the extremely hot loops in corona simultaneously.
%While the coronal activities are dominated by the magnetic field, thus the magnetogram gained from Helioseismic and Magnetic Imager \citep[HMI,][]{scherrer2012} are also included in our study.

For analyzing the 3D coronal magnetic field, the vector magnetogram of photosphere is extrapolated based on the nonlinear force-free field (NLFFF) model. We note that the photospheric magnetic field is generally not force-free, and the force-free field is a reasonable approximation for only the low $\beta$ environment in the corona, where the core structures taken part in the flare process and eruptions are mainly embedded in. The vector magnetogram is gained from the Space-weather HMI Active Region Patch \citep[SHARP,][]{bobra2014} dataset and the NLFFF extrapolation is realized by an MHD relaxation method, namely, the CESE--MHD--NLFFF code developed by \citet{jiang2013}. This method seeks the equilibrium magnetic field in the zero-$\beta$ environment with friction force via solving a set of modified MHD equations. An advanced conversation-element/solution element (CESE) space-time scheme on a nonuniform grid and parallel computing are used in this method \citep{jiang2010}. It was tested well by different benchmarks~\citep{Jiang2012,Jiang2016}, such as the analytic force-free solutions by \citet{low1990} and numerical MFR models \citep{titov1999,van2004}. Previous works show that magnetic configurations extrapolated from SHARPs show good correspondence with the observable features, such as coronal loops, filaments and sigmoids \citep{jiang2013,jiang2014}. The extrapolation quality of the code has been well examined in \citet{jiang2013}.
In particular, here the extrapolated field has a high level of solenoidality, as the mean divergence error is $\sim 5\times 10^{-4}$ as measured by $f_i = |\nabla\cdot \mathbf{B}|/(6B/\Delta x)$ {(where $\Delta x$ is the grid resolution) for all the extrapolations \citep{jiang2013}}.
To study the evolution of the MFR, we carried out extrapolation from 00:00 UT to 11:48 UT, (i.e., from the begin of September 6 to the time immediately prior to the X9.3 flare) with cadence of 12 min. Note that there is an 2-hour data gap of HMI between 6:00 UT to 8:00 UT. Before extrapolation, we have carefully selected the magentograms as we found that at certain times, the SHARP magnetogram shows clear artifacts (which are some stripe-like features in horizontal field seen near the upper sunspot) which seems to be unreasonable and are thus excluded in our extrapolations.

For analyzing details of the MFR structure, %magnetic topology as well as the torus instability,  %Commonly, some of the magnetic structures are more likely to cause the violent activities. In order to pick them up,
we compute different parameters for the extrapolated magnetic field, including the magnetic twist number $T_{w}$, the decay index $n$, the self magnetic helicity $H$ of the MFR and the magnetic energy. The twist number describes the winding turn of two infinitesimally close field lines using the follow equation \citep{liu2016}:
\begin{equation}
T_{w}=\int_{L}\frac{\left(\nabla\times \mathbf{B}\right)\cdot \mathbf{B}}{4\pi B^2}dl.
\end{equation}
%Previous works proposed that when the twist number is high enough, the kink instability will be triggered \citep{mikic1990}.
Here the integral is taken along magnetic field line from one footpoint to the other one on the photosphere {and $L$ is the length of the given field line}, so here the twist number is computed only for closed field lines. The decay index $n$, which describe the decay speed of the strapping field strength $B_{p}$, whose Lorentz force will constrain the expansion of underneath magnetic structures, with the distance $h$ from the bottom surface, can be calculated by $n=-\partial\left(logB_{p}\right)/\partial\left(logh\right)$. Note that here $B_{p}$ is approximated from potential field model and only the component perpendicular to the path direction along which we compute $n$ is used as $B_p$.  Some theoretical studies indicate that the torus instability will be triggered when the axis of MFR is above the threshold $n > 1.5$ \citep{bateman1978,kliem2006}.
{The magnetic helicity $H$ can quantify the degree of the shear or twist of magnetic field \citep{demoulin2006}. It can be divided into two parts, one is self helicity $H_s$ and the other is mutual helicity $H_m$. If considering that an MFR is approximated by a set of flux tubes, we can use the self helicity to quantify the sum of the flux tubes' individual twists and writhes, and the mutual helicity to quantify the linkage and knotting between different flux tubes \citep{guo2017}. In order to clarify their usage, \citet{demoulin2006} concluded that the mutual helicity can be used as a substitute of the total magnetic helicity, since the self helicity is only $1/N$ of the mutual helicity, where $N$ is the number of flux tubes involved in the computed magnetic configuration. While \citet{guo2017} found, under the assumption that an MFR consists of finite number of flux tubes, the self helicity is still useful. For example, a preliminary work of them \citep{yang2016} shows that although the order of magnitude of self helicity is lower than that of mutual helicity, their evolution trend is similar. It means, if we only concern the evolution trend of helicity, using the self helicity is more convenient since its calculation is simple. Thus here we only compute the self helicity of the MFR following the method of~\citep{Guo2013,guo2017} by using equation $H_s = \sum_{i=1}^NH_{si} = \sum_{i=1}^NT_{wi}F^{2}_{i}$, where $N$ is the number of magnetic flux tubes whose twist number is higher than 1, $H_{si}$ is the self helicity of $i$th magnetic tube, and $T_{wi}$ and $F_{i}$ is its twist number and flux. Note that we do not consider the influence of writhe since there is no significant writhe can be seen in the MFR.}
%Here we only compute the self helicity of the MFR by using the equation $H_{s}=T_{w}F^{2}$. Where $H_{s}$ is self helicity, $T_{w}$ is twist number of the twisted field line and $F$ is its flux.

\begin{figure*}
\centering
\includegraphics[width=\textwidth]{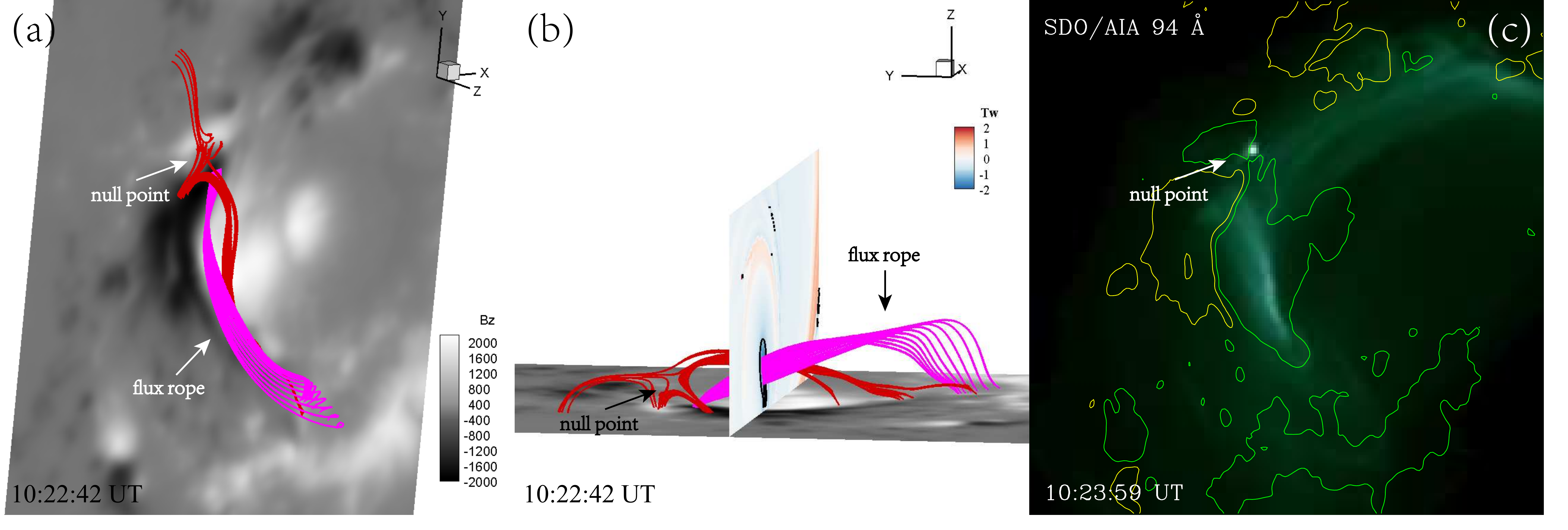}
\caption{The extrapolated magnetic configuration of the core structure. In panel (a) and (b), the red magnetic lines are the fan-spin structure of null point and the purple lines are the magnetic flux rope, and they are shown with different 3D view angles. Panel (a) is the magnetic field lines project onto the plane-of-sky, thus can be compared with the SDO/AIA  94 \AA\ image shown in panel (c). The green (yellow) contours in panel (c) denote line-of-sight magnetic field of 500~G ($-500$~G). Panel (b) is side view. A longitudinal slice is selected to show the twist distribution and it is also the selected slice shown in Figure~\ref{fig3}. }
\label{fig1}
\end{figure*}

\section{Results} \label{res}

% 这一段的信息和Introduction信息重复，可以删除或者补充到Introduction 里面。Results 部分要直接介绍结果，无需背景介绍。As we mentioned in section \ref{sec:intro}, two large flares occurred in AR 12673 at Sept. 6 2017. These two flares are both X-class flares and interestingly, one was confined and the other one was accompanied with a huge CME. Since both of them located at the same position, i. e., the PIL of two largest sunspot in this AR, thus the magnetic configurations above the PIL is dominate the evolution and eruption of these two flares. In previous works, many authors indicated the MFR above the PIL play a most important role in triggering flare or eruption \citep{yan2018,liu2019,zou2019}.
To trace the evolution of the MFR structure, we first exhibit the basic configuration of the core magnetic field, and then identify the MFR using the 3D distribution of the twist number. To give more view of this MFR, cross section of the MFR is also analyzed. Figure~\ref{fig1} shows the basic 3D configuration of the pre-flare coronal magnetic field, which contains an MFR (shown by the purple lines) and its overlying magnetic field (the red lines). The overlying field consists of a spine-fan structure of a magnetic null point located at north-east of the MFR. This null point was first found in the analysis of \citep{zou2019} and was considered as the trigger origin of the X2.2 flare. The MFR, whose north leg located in north part of east spot and south foot located in south part of west spot, is the core structure responsible for these two flares \citep{jiang2018,zou2019}.

\begin{figure*}
\centering
\includegraphics[width=0.8\textwidth]{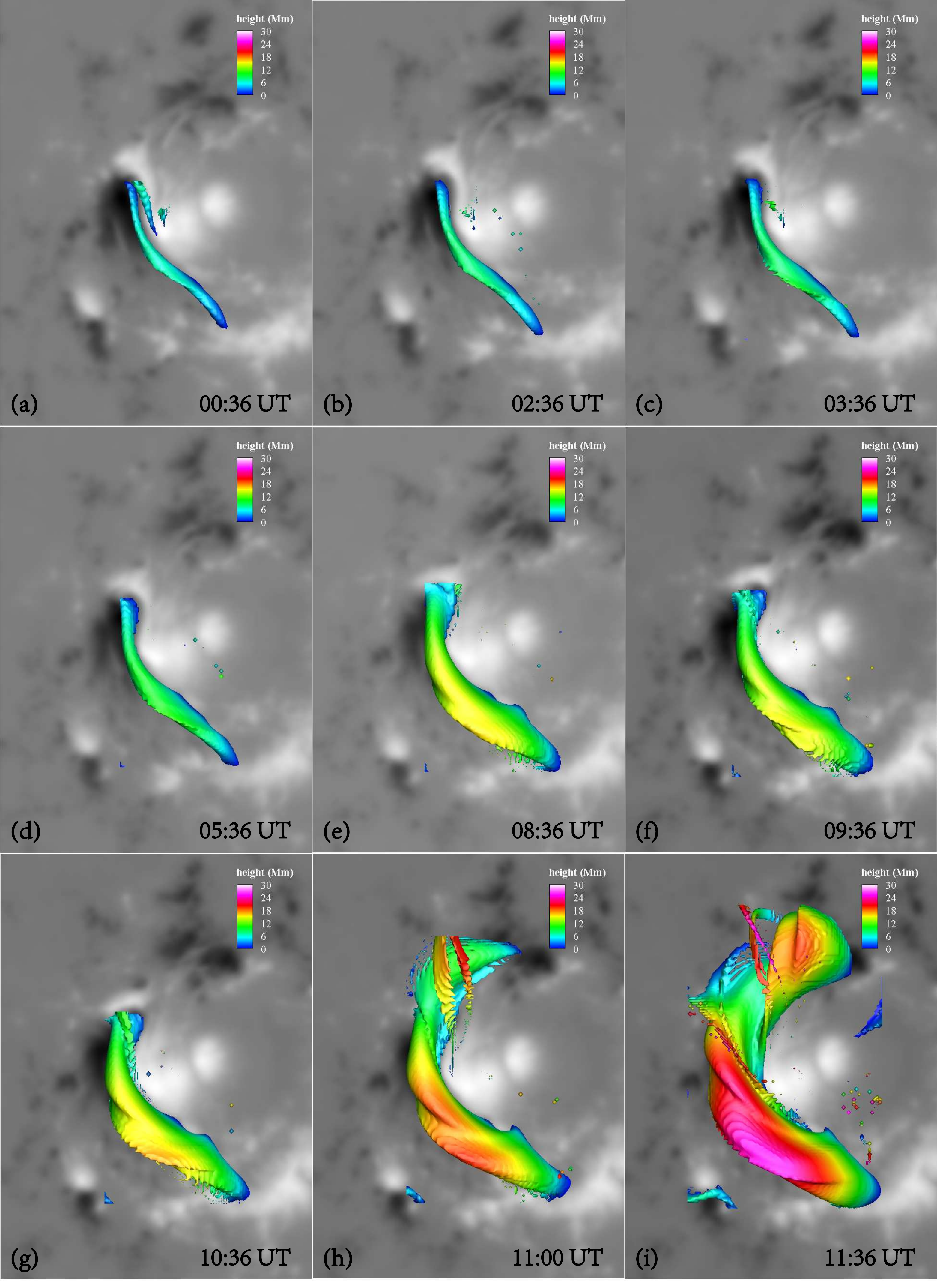}
\caption{The evolution of the 3D structure of MFR from 00:36 to 11:36. The 3D colored structures are shown by the iso-surfaces with {$|T_w|=1$}, and the color denotes the height of different parts. The background gray images show vertical component $B_z$ of magnetic field on the photosphere.}
\label{fig2}
\end{figure*}

\begin{figure*}
\centering
\includegraphics[width=\textwidth]{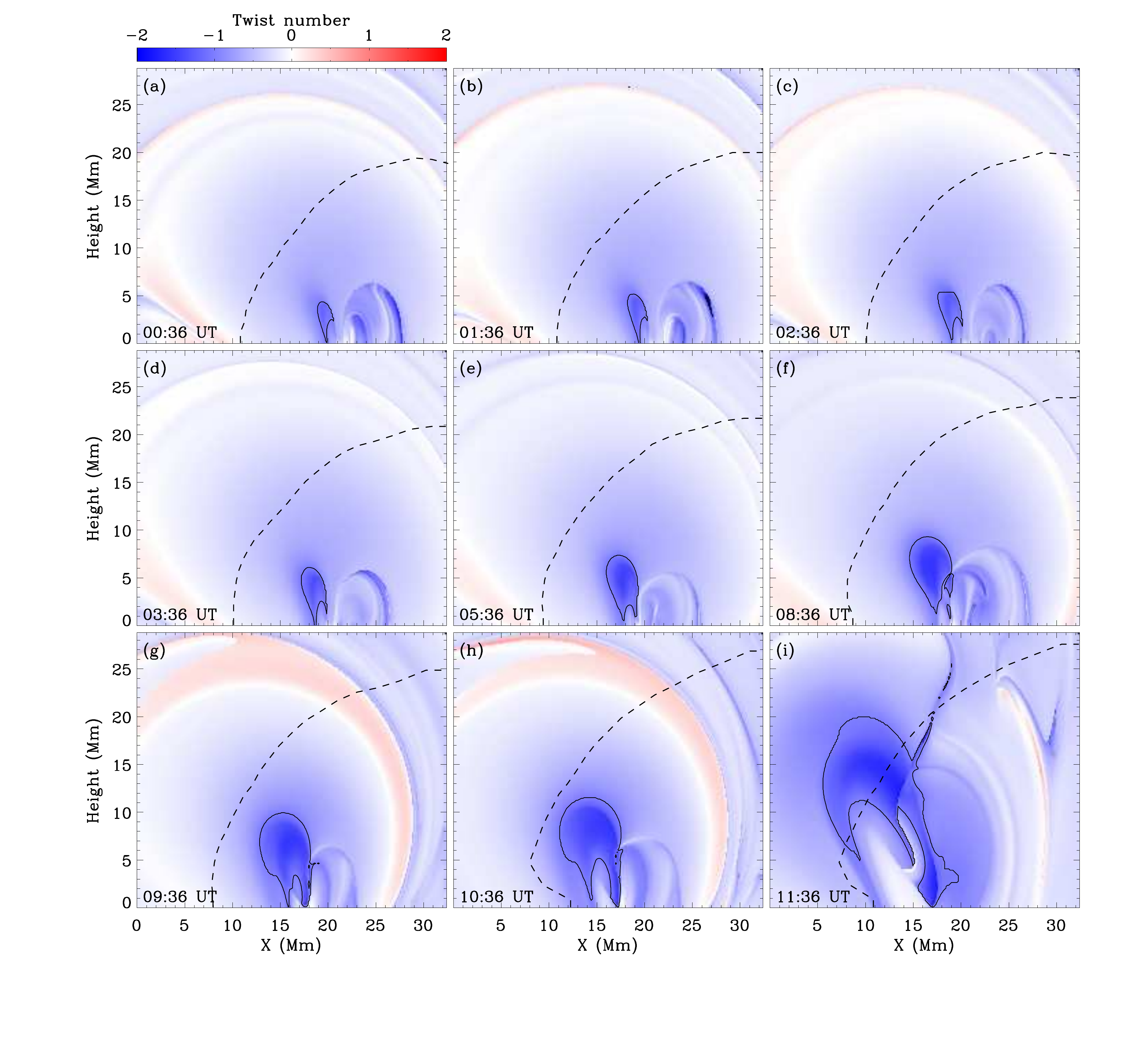}
\caption{Evolution of magnetic twist number distribution on the vertical cross section (whose location is shown in Figure~\ref{fig1}) from 00:36~UT to 11:36~UT. {The black solid lines show the limits of the the MFR, i.e., the contour lines with $T_w=-1$. The black dashed lines show threshold of torus instability, i.e., the line on which the decay index $n=1.5$. } }
\label{fig3}
\end{figure*}

\begin{figure*}
\centering
\includegraphics[width=0.8\textwidth]{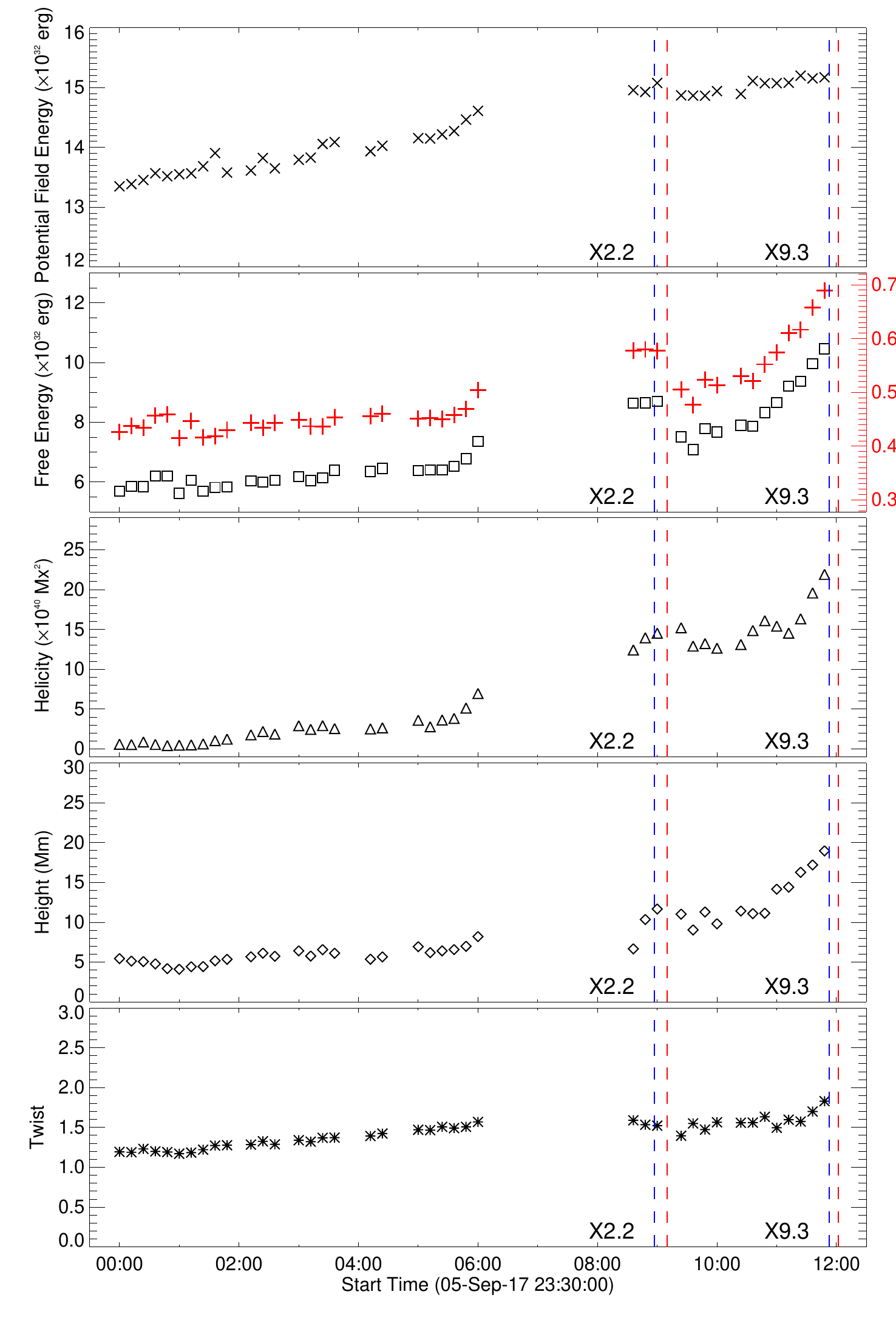}
\caption{Time evolution of potential field energy and free magnetic energy, helicity, twist, height of MFR axis, and the {absolute value} of peak twist number in the MFR. The axis height is evaluated from the axis of MFR in selected slice of Figure~\ref{fig3}. The potential field energy and free magnetic energy was quantified from the whole area we extrapolated, {and the ratio of the free energy to the potential energy is also plotted in the second panel}. The flare start time and peak time are indicated in the figure using blue and red dashed lines respectively.
}
\label{fig4}
\end{figure*}

In Figure~\ref{fig2}, we present the evolution of 3D structure of the MFR. Here the MFR is defined as a coherent group of magnetic field lines with unsigned twist number above 1~\citep{liu2016b, DuanA2019}, thus the whole structure of the MFR can be shown by the iso-surfaces of magnetic twist number $|T_w|=1$ and the high twisted flux is wrapped by this surface. In Figure~\ref{fig3} we show the magnetic twist distribution on a vertical central cross section of the volume (the position of this slice is shown in Figure~\ref{fig1}). Furthermore, in Figure~\ref{fig4} (the last panel), the evolution of the largest magnetic twist number in the MFR is shown. The reconstruction reproduced clearly the growth of the MFR from the beginning of the day until the X9.3 flare. Initially, the MFR is rather thin and the twist number is mostly below 1.5. It continuously expands in volume and rises in height as driving by the photosphere motion of the sunspots. Until the first flare (8:57~UT), the MFR grows rather slowly. While after the first flare, it expands much faster until the second flare, reaching a maximum height of $\sim 30$~Mm and a volume much larger than that before the first flare (comparing panels (e) and (i) in Figure~\ref{fig2}), as more and more flux joins in the MFR. Furthermore, the MFR extends to the polarity in the northwest of the AR, forming a C shape. On the other hand, the peak twist number of the MFR increases slightly from 1.5 to $\sim 1.8$ in the duration between the two flares.

In our prior study \citep{zou2019}, it is indicated that the heights of MFR before these two flares make the key difference to determine the eruptiveness of the flare, i.e., whether the axis of MFR is above or below the threshold of torus instability. To give prominence to the relationship between MFR height and torus-instability threshold during the time evolution, we also plot the contour line of critical value of decay index $n=1.5$ on the central cross section, as shown by the black solid lines in each panels of Figure~\ref{fig3}. Clearly, the twisted magnetic flux is located below the threshold line before the X9.3 flare, while immediately before it, a major part of twisted flux is above the threshold. Moreover, the torus-instability threshold line shows small variation during the evolution, since it is the rising of the MFR that makes it to approach the unstable regime until the X9.3 flare. This is in well agreement with the scenario that the MFR was approaching the threshold during its evolution and finally erupted once it go through the threshold.

In addition to the peak twist number, Figure~\ref{fig4} also shows time evolution of some other parameters of the magnetic field, including magnetic energies, self helicity of the MFR and the height of the MFR axis. Here the axis of the MFR is defined to be the field line in the rope that possesses the peak value of magnetic twist number. It can be seen that before the X2.2 flare, both the magnetic free energy and helicity increases monotonously but rather slowly. During the X2.2 flare, the magnetic free energy drops stepwisely, with a release of $\sim 1.3\times 10^{32}$~erg, while there is nearly no {decrease} of magnetic helicity {in the extrapolation volume}, which is consistent with the nature of confined flare because {otherwise eruption will take part of helicity away from the active region}. After then, the free energy increases much faster until the X9.3 flare, gaining a value of $\sim 3.5\times 10^{32}$~erg, and the helicity also gains $\sim 7\times 10^{40}$~Mx$^2$. It is interesting to compare the relative measure of the non-potentiality, e.g., ratio of the free energy to potential energy, with that shown in \citet{sun2015}, who suggests that such relative ratio might be more directly correlated to the eruptiveness or confinement of the flare than other global quantities. Here we can see that prior to the first confined flare and the second eruptive one, the energy ratio is always higher (0.58 for the X2.2 flare, and 0.69 for the X9.3 flare) than that of the eruptive ARs studied in~\citet{sun2015}. Thus, by such relative ratio we still cannot discriminate the types of the flare in the studied AR. The helicity and its increasing rate are significantly high if compared with those of other ARs in pre-flare phase~\citep{Guo2013,yang2016}. {This is due to the high flux content of the MFR here, as the increase of the MFR's self helicity is consistent with the expansion of the MFR and increase of its flux.}
 %And as a comparison, the helicity calculated by \citet{vema2019} and \citet{moraitis2019} are significantly higher than previous works either.}
%This suggests that in the short time interval between the two flares, there is a fast injection of magnetic energy but without much magnetic twist resulted.
Meanwhile, we note that the potential field energy is relatively stable, almost without changing between the two flares when the free energy increased rapidly, which suggests that the increasing free energy is not due to the emergence of flux from underneath the photosphere but accumulated by the rotating and shearing motion of photosphere~\citep{yan2018}. The MFR axis height rises slowly before the X2.2 flare, from 5~Mm to 10~Mm, and after a transient decline across the X2.2 flare, it rises again and much more rapidly, from 10~Mm to 20~Mm. Such evolution pattern is nicely in correspondence with the evolution of the free magnetic energy. The absolute value of peak magnetic twist number shows systematic and slow increase from 1.2 to 1.8 without much transient changes in the whole duration from the beginning of the day to the X9.3 flare. Although some theoretical studies suggest that the kink instability could occur if the twist number is larger than $\sim 1.25$~\citep{Hood1981}, here the MFR did not erupt even its twist number reaches above 1.5 after 6:00~UT, and furthermore no kink motions can be seen during whole observations. It seems to be common in many investigations~\citep{Guo2013,liu2016b}, i.e., the MFRs do not significant kinked even their twist number is higher than 1.5, and a recently statistical study suggests that the lower limit of kink instability could be $|T_w|=2$~\citep{DuanA2019}.

%The surrounding environment seems contribute more in free energy increase. textbf{While the rapid raise of axis was half an hour before the twist increase. It also suggests that twist increase may not the main reason why MFR raise.}

\begin{figure*}
\centering
\includegraphics[width=\textwidth]{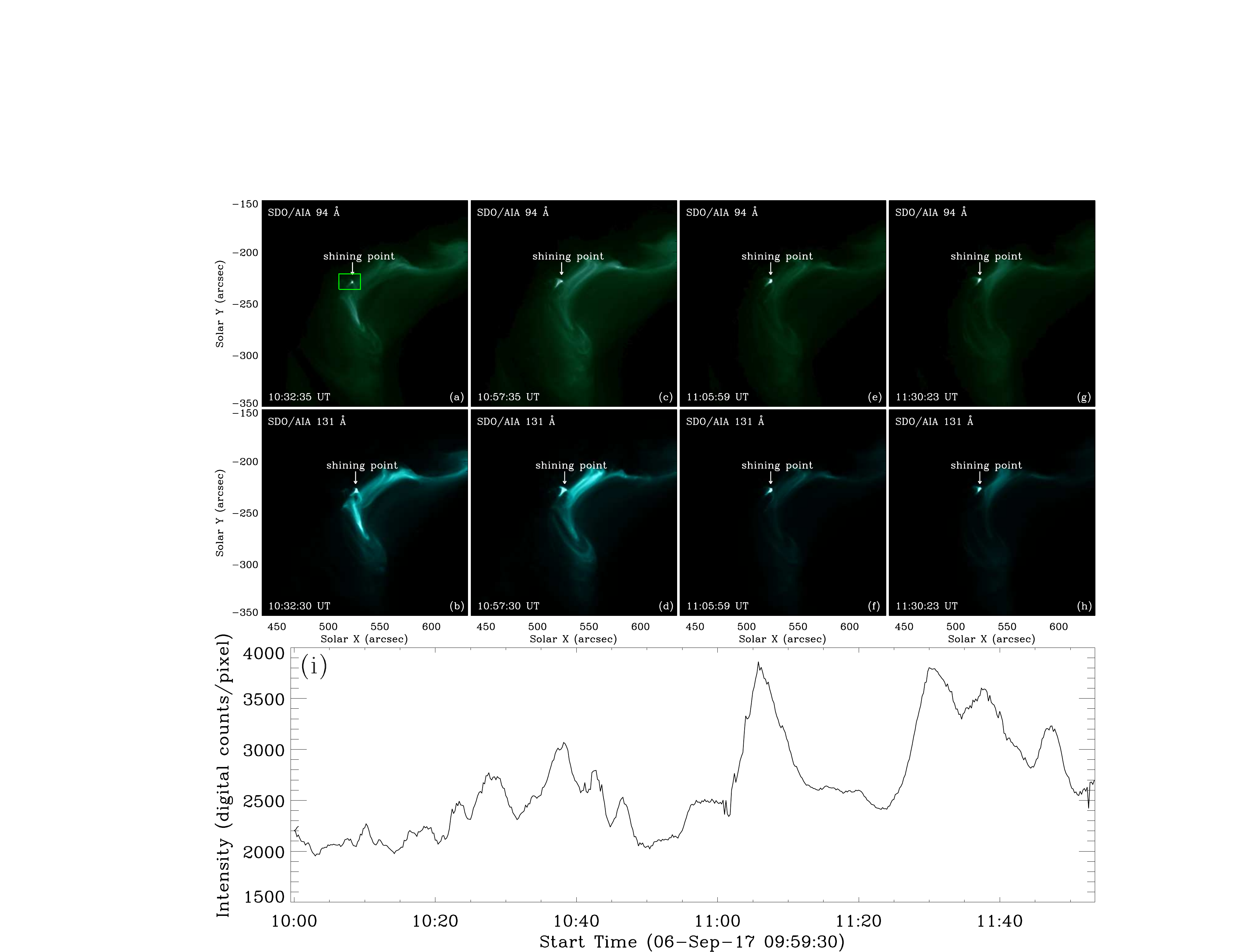}
\caption{Panel (a) to (h) are the shining null point seen in 94 \AA\ and 131 \AA\ images at different time respectively. Panel (i) shows the 94 \AA\ light curve of bright features where are indicated in panel (a) using a green box. Note that the bright features are selected via an intensity threshold $I\geq1500$.}
\label{fig5}
\end{figure*}

From the above analysis, it is suggested that the fast rising of the MFR in altitude (which occurred between the two flares) rather than the increase of the magnetic twist of the rope plays the key role in leading to the eruption.
%Therefore, in order to understand why the MFR will erupt and trigger the flare, we should understand the reason of the MFR's upward motion.
So, what is the key factor that leads to the fast rising of the MFR after the X2.2 flare?
%Generally, there are two factors can be taken into consideration. One is the enhancement of the internal stress of MFR, such as the increasing of twist. The other one is weaken of constrained force, similar to the breakout model. Actually, as we shown in Figure~\ref{fig4}, before the X2.2 flare, the MFR surely rising with the enhancement of twist or helicity slowly. But after the X2.2 flare, the twist or helicity become stable and their enhancement was occurred soon before the X9.3 flare.  However, the raise of MFR after X2.2 flare occurred before the twist or helicity was rapidly enhanced before X9.3 flare. It means, the inner force increasing is not the main reason why the MFR start rising. Thus the weakening of overlying field is more probably. In order to understand how the overlying magnetic field was weakened,
In order to answer this question, we rechecked the AIA observation in hot channels and found a very interesting phenomenon. As mentioned before, on the northeast of MFR, there was a cusp structure, which is recognized as a null point whose fan extends overlying the northern part of the MFR. After the X2.2 flare, strong emission is seen recurrently at this null point structure in hot emission lines of SDO/AIA 94 and 131~\AA, as shown in Figure~\ref{fig5}, and the strong intensity even generated diffraction fringe often seen in flares (see in the animation accompanied with Figure~\ref{fig5}). Such a null-point topology as well as the enhanced heating strongly indicate that magnetic reconnection was continuously taking place there. The reconnection will weaken the overlying field and help the MFR rise, and vice verse, the raise of MFR will push more flux to the null point for more subsequent reconnection. As a result the reconnection is sustained, along with the fast rising of the MFR until the it runs into the TI threshold.

\section{Discussion and conclusion} \label{dac}

In this paper, we have investigated an MFR in AR~12673, which evolved from the beginning of 2017 September 6, experienced an X2.2 confined flare, and finally erupted to an X9.3 eruptive flare. The main interest for us is how the MFR is formed before the first flare and evolved between the two flares until triggering the second flare. Actually, previous works have discussed some aspects of evolution of this AR associated with these two flares. For instance, it has been mentioned the shearing and rotating motions of sunspots on photosphere are the key point in building up the MFR to eruption \citep[e.g.,][]{romano2018, yan2018, vema2019}.
%also mentioned that the vortex motion of the negative spot's north part (also the north foot of MFR) build up the core structure and lead to eruption.
While these studies show generally the energization of the coronal field comes from the surface motions, it is still not clear how the MFR evolved in 3D. To clarify this, we constructed a time sequence of NLFFF reconstructions matching the cadence of the HMI vector magnetograms to follow the quasi-static evolution of the coronal magnetic field, from which a full 3D information of the MFR evolution can be derived.

%These are consist with our results based on NLFFF extrapolation, i.e., the vortex motion of footpoint inject helicity and twists into the magnetic arcades and build up an MFR and this MFR became the core structure of eruption.

It is informative to compare our results with a recent similar study by \citet{liu2019} for the same AR but with a different extrapolation method. In their study, the extrapolation show similarly a MFR structure before the flares. However, their extrapolated MFR shows no significant changes, e.g., raise in height, between the two flares, and the MFR is far below the torus-instability threshold even near the X9.3 flare. As such, they proposed that the MFR system is enhanced during the X2.2 flare and the final eruptive flare is triggered by the internal reconnection between the twisted magnetic flux, which is similar to the so-called "domino-effect" scenario \citep{zuccarello2009}. While in our extrapolation, the MFR shows a significant change between the two flares, i.e., after the X2.2 flare, it rises continuously and reaches the threshold of TI immediately before the X9.3 flare. Of course, we cannot exactly know which extrapolation is the true structure of AR 12673. But in our extrapolation, the results can clearly indicate the main difference between two flares, namely the triggering of torus instability. Furthermore, our prior analysis \citep{zou2019} suggested that due to the strong confinement of the overlying flux, reconnection alone, either internal tether-cutting or external breakout reconnection, seems not able to trigger the eruption in this AR. Such conclusion is enhanced in this paper, as there appears to be continuous reconnection at the null point aside of the MFR but the eruption can only be  trigger until the MFR rises across the torus-instability threshold.
%supported by the simulation of \citet{aulanier2010} which suggested that only reconnection can hardly erupt a CME. %Thus we think our analyse is more reasonable.

Nevertheless, in terms of the raise of MFR, we have the similar conclusion with \citet{liu2019} that the X2.2 flare helped building the MFR to the final eruption. %They think in terms of two close activities, people should pay more attention in the previous activity, since it may activate the magnetic structures.
This is because the observation of AIA shows that the recurrent reconnection in the null point is triggered after the X2.2 flare. It means this null-point became active after the X2.2 flare somehow. \citet{zou2019} shows that the null point reconnection is the first reconnection episode in the two-step reconnection scenario in the X2.2 flare, and it was most probably triggered by the upward motion and expansion of the underneath MFR, i.e., at the start of first flare, the null point was activated by the MFR.
%Actually, we also notice that after the X2.2 flare, while the height of MFR axis is fluctuating, but the diameter of MFR is expanding (see panel g and h in Figure~\ref{fig2}).
Thereby the expansion can push the null point and force it reconnect recurrently. Eventually, the MFR reached the threshold and eruption occurred. So in this scenario, the eruption of the X9.3 flare is also due to that the overlying structures were activated by the preceding confined flare.

In summary, we suggest a scenario of the evolution of the MFR leading to the two major flares. Initially, the sunspot rotation on the photosphere build up an MFR, with its magnetic flux, twist as well as height gradually increasing. When it is high enough, it can push the overlying field and force the null point reconnection of the overlying field. The null point was activated and lead to the onset of the X2.2 flare. The recurrent reconnection at the null point continually weakens the overlying field and allow further, and faster upward expansion of the MFR. Eventually, the MFR went across the threshold of torus instability and erupted. This scenario is essentially similar to the breakout model, i.e., the reconnection weakens the overlying magnetic field and allows the stressed magnetic flux underneath to expand until its eruption \citep{antiochos1999}. However, the process in the studied event is more gentle and last a longer time period, and the eruption trigger is the torus instability.

%In this scenario, both reconnection and instability contribute to the final eruption, they can not isolate from each other. Also, an eruption can be leaded by many complicate process, not only the shearing/converging flows, flux emerge or filament oscillation, but also some tiny shinning nearby. Since they can slightly change the magnetic configuration and eventually construct an eruptive structure.

\acknowledgments
This work is supported by the National Natural Science Foundation of China (NSFC 11903011, 41822404, 41731067, 41574170, 41531073), the Fundamental Research Funds for the Central Universities (Grant No.HIT.BRETIV.201901), and Shenzhen Technology Project JCYJ20180306171748011. P.Z. also acknowledges the support by China Postdoctoral Science Foundation (2018M641812). Data from observations are courtesy of NASA {SDO}/AIA and the HMI science teams. C.W.J. thanks ISSI-BJ for supporting him to attending the team meeting led by J.C.Vial and P.F.Chen.

\end{document}